\documentclass[aps,prl,nofootinbib,showpacs,twocolumn]{revtex4}
\usepackage{amssymb,latexsym,mathrsfs}
\usepackage{amsfonts}
\usepackage{amsmath}
\usepackage{graphicx}
\newcommand{\beq}{\begin{equation}}
\newcommand{\eeq}{\end{equation}}
\def\bea{\begin{eqnarray}}
\def\eea{\end{eqnarray}}
\def\ba{\begin{array}}
\def\ea{\end{array}}

\begin{document}
\title{Two-dimensional random walk in a  bounded domain}
\author{Mahashweta Basu}
\email[E-mail address: ]{mahashweta.basu@saha.ac.in}
\author{P. K. Mohanty}
\affiliation{Theoretical Condensed Matter Physics Division,\\ Saha Institute of Nuclear Physics,
1/AF Bidhan Nagar, Kolkata, 700064 India.}
\date{\today}

\begin{abstract}
 In a recent Letter Ciftci and Cakmak [Eur. Phys. Lett. {\bf 87}, 60003 (2009)] 
showed that  the two dimensional random walk    in a bounded 
domain, where walkers which cross the boundary return to a base curve 
near origin with deterministic rules, can produce regular patterns.  
Our numerical calculations suggest that  
the cumulative probability distribution function of  the 
returning walkers along the  base curve  is a Devil's staircase, 
which can be explained  from the mapping of these walks  to a non-linear 
stochastic map. The non-trivial  probability distribution function(PDF) is a 
universal feature of CCRW characterized by  the  fractal dimension 
$d=1.75(0)$ of the PDF  bounding curve. 
\end{abstract}
\maketitle
Diffusion is a basic physical process \cite{Diffusion} that moves 
substances  randomly from the high  concentration regimes to the low ones.  
A  simple mathematical model of this  random  Brownian motion   
\cite{Brown, Einstein} is well described  by a random walk \cite{Pearson}  where 
the walker takes unit steps successively  in an arbitrary direction. 
Several variations of the  random walk with different boundary conditions \cite{Feller}
has been studied  in $d$-dimensions.  Such walks can also be defined on a lattice
where,  starting from origin, the walker moves randomly to one of its neighbour.  
The individual random walks  are known to become scale invariant after a large number 
of steps and their radial distances from origin  follow a normal distribution.
Being simple, the models of random walk and some of their 
variations, have found applications in  several branches of science. 
Reaction-diffusion systems, percolation,  network dynamics and  
stock fluctuations \cite{Sethna} are few to  name.

Different boundary conditions \cite{Feller}  are known  to have
strikingly  unusual effects on the longtime behaviour of the random 
walks. Recently  Ciftci and Cakmak \cite{Ciftci}  studied  the two
dimensional random walk (2DRW) in a bounded region, where the 
walker deterministically returns to a  pre-defined base curve  
near origin after crossing the domain boundary. For example, in one 
particular case, the walker chooses  a new coordinate 
$(x \to x / \sqrt{x^2 +y^2}, y \to y/ \sqrt{x^2 +y^2})$ after crossing 
the boundary ($|x|< 6$ or $|y|<2$)  and  starts a fresh walk.  
Interestingly these walks, hereafter named as   Ciftci and Cakmak
random walk (CCRW)  produce   regular patterns.

In this  Letter we show that the patterns are simple repetition of
the base curves with centers placed at  all the points 
$(i,j)$  within the domain where sum of these integers 
$(i+j)$ is even.
The cumulative distribution 
of returning walkers along the base curve  is found to be a Devil's 
staircase (DS) \cite{DS}, which could be explained from the mapping of CCRW to a 
stochastic non-linear map. These DS structures, which are  generic 
features of CCRW, can be characterized by  the fractal dimension 
$d$ of curve which bounds the  corresponding  PDF.  Numerical 
calculations suggest that $d= 1.75(0)$ is universal.

\begin{figure}[h]
\centering
\includegraphics[width=6cm]{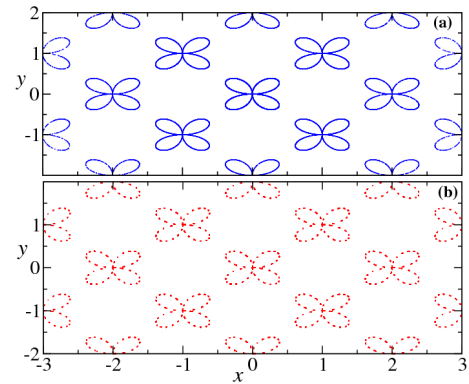}
\caption{Patterns : (a) uncorrelated  random walk  starting 
from the base curve (\ref{eq:flower}), (b) corresponding CCRW after
$10^5$ steps.}
\label{fig:flower}
\end{figure}
 
In CCRW, the walker initiate a  discrete time walk on a two dimensional 
square lattice  starting from origin,  by  taking  a random
unit step $\pm 1$   both in $x$ and $y$ direction. Thus  in each time step its 
coordinate changes from $(x,y)$ to $(x+\sigma^x, y + \sigma^y)$, 
where  $\sigma_{x,y}=\pm 1$ chosen  randomly. The walk is confined in a bounded 
domain, usually taken as rectangle $|x|<b_x$  and $|y|<b_y$. 
If the walker crosses the boundary , $e.g.$ when  $|x|>b_x$ or $|y| >b_y$,
it returns  immediately to  a  new coordinate $(f(\theta), g(\theta))$, where 
$\theta = tan^{-1} (y/x)$.  The new co-ordinates define a curve 
\beq 
x= f(\theta) ;~~ y = g(\theta)
\eeq
parametrized by $\theta$, which will be referred to as the base curve (BC).
Case-I of \cite{Ciftci} corresponds to $f(\theta) = \sin \theta$ and 
$g(\theta) = \cos\theta$ which  generate regular patterns in a bounded domain, 
compared to simple two dimensional random walk which visit only a set of integer 
points $\{\vec v \}= (x=i,y=j)$  where  $(i+j)$ is even.

The walker in CCRW eventually returns to the base curve and starts a 
fresh walk from there.  Thus, it is natural to expect that 
the patterns of CCRW can also be generated  by  a set  of uncorrelated random walk 
(URW) of a large number of walkers starting from the base curve. In URW, however, 
one must use a absorbing boundary condition  that the walk terminates  
when the walker crosses the boundary.  Starting from any arbitrary point $\vec r$ 
on the base curve each walker visits only  the  points  $\{\vec r + \vec v \}$
Hence, URW produces  a pattern where the base curve is 
shifted by the vectors $\{\vec v \}$. This pattern is independent of the 
initial distribution  of walkers on the base curve.

To show that  URW can generate any desired pattern we  take  an 
example,
\bea
 {\rm Boundary} &:& |x|\le 3 ~~{\rm and } ~~|y|\le 2\cr
 {\rm Base ~ curve} &:& f(\theta)= a\sin(\theta) \cos^2(\theta) \cr  
 &~& g(\theta)=a\sin^2(\theta) \cos(\theta).
\label{eq:flower} 
\eea 
 
Let the distribution of  walkers $P(\theta)$ along the base curve at an angle $\theta$ is   
be uniform [$i.e$ $\theta$ is a random number uniform in $(0,2\pi)$]. The patterns  
generated by this URW with $a=1$ is shown in Fig. \ref{fig:flower}. 
In this  pattern the repeating base curves are non-overlapping as their lobes  are 
bounded within a circle of radius $1/{\sqrt 2}$. One can, in fact, generate  more 
complicated and overlapping patterns either by choosing  $a> {\sqrt 2}$ or by taking
different BCs.  
  
\begin{figure}[h]
 \centering
\includegraphics[width=8cm,bb=0 200 792 612]{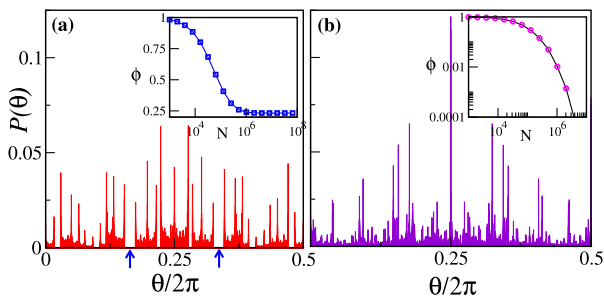}
\caption{Distributions of returning walkers $P(\theta)=P(-\theta)$ 
along the BC  (\ref{eq:flower}), and a unit circle are shown 
in (a) and (b) respectively. Clearly, $P(\theta)$ is 
symmetric about $\theta=\pi/2$.  The arrows in (a) point to some forbidden 
regions where $P(\theta)=0$. Here, PDFs are calculated with 
$\Delta\theta= 2\pi 10^{-5}$ and  $N=10^8$.  The insets of (a) and (b)
show variation of $\phi$ with $N$ for respective BCs ($\epsilon=10^{-4}$).}
\label{fig:pdf}
\end{figure}

Patterns similar to Fig. 1(a) can also be produced by CCRW.  
The walker in this case initiates the walk 
from origin and immediately after crossing the  boundary it  returns 
to the base curve (\ref{eq:flower}).  The return is deterministic as 
$\theta = \tan^{-1} (y/x)$ in Eq. (\ref{eq:flower}) depends on the 
final co-ordinates  $(x,y)$ of the walker.
Corresponding pattern is shown  in Fig 1(b). Unlike Fig 1(a), here, 
the patterns appear discontinuous, which indicates that certain 
regions along the BC are never visited. This is also confirmed from 
the numerical results  of the distribution of 
returning walkers $P(\theta)$ along the BC in the range $\theta=\in (0,\pi)$. 
Fig. \ref{fig:pdf} shows the distribution $P(\theta)$, where 
some of the forbidden regions [$P(\theta)=0$] are marked with an arrows. 
A quantitative measure of the forbidden regions is  
\begin{equation}
\phi=  \int_0^{2\pi} \Theta(\epsilon-P(\theta)) d\theta  ,
\end{equation}
where $\epsilon \simeq 0 $ is a  pre-determined 
positive number and $\Theta(x)$ is the Heaviside step function.
In fact  $\phi$ saturates to $\phi_s=0.232(4)$ as the number of 
time steps $N$ increases [Inset of Fig. \ref{fig:pdf}(a)]. A non-zero  $\phi_s$
is not a typical characteristic of CCRW. When the BC is a circle
(Case -I of \cite{Ciftci}), we find that  the  walker returns to 
all values  of $\theta $ with finite probability [as shown 
in  Fig. 2(b)] and thus $\phi_s\to 0$ [Inset of Fig.  2(b)].

A unique feature  of CCRW that emerges from  the distribution  of returning 
walkers shown in   Fig. \ref{fig:pdf} is that this probability measure may not 
be represented in any functional form. To get a functional form, namely the PDF, 
one  needs to count the fraction of walkers $P(\theta |\Delta\theta)$ that comes back to 
a bin of size $\Delta \theta$ about $\theta$ and then take the limit
\cite{Feller} 
\begin{equation} 
 P(\theta)=\lim_{\Delta\theta\to 0} P(\theta|\Delta\theta). 
\end{equation}
The PDF is well defined only when above limit exists. In Fig. 3(a) 
we have plotted  $P(\theta|\Delta\theta)$  against $\Delta \theta$ 
for  CCRW with BC (\ref{eq:flower}) and for two different values  
of $\theta = 0.7, 0.4$. It is evident that  the limit $\Delta \theta \to 0$ 
does not exist. Figure 3(b) there shows the same  for
Case-I of \cite{Ciftci} and for  $\theta = 0.7, 0.3$. 
\vspace*{.5 cm}

\begin{figure}[t]
\centering
\includegraphics[width=6cm,bb=50 80 790 552]{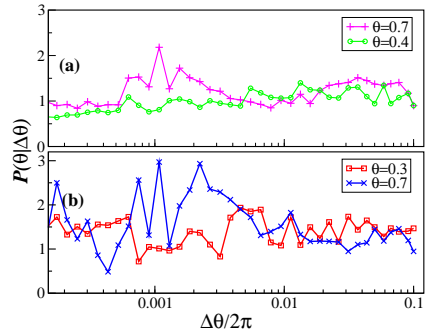} 
\caption{ $P(\theta|\Delta\theta)$  does not approach any definite  limit 
as $\Delta\theta \to 0$. (a) and  (b)correspond to CCRW with 
base curve (\ref{eq:flower}) and a unit circle respectively.}
\label{fig:limit}
\end{figure}

It is natural to ask, if the cumulative distribution  defined by 
$P(\theta >\phi) = \int_0^\phi P(\theta) d\theta$ is a well defined 
function. Fig. 4 shows that  $P(\theta >\phi)$  against $\phi$  
for the base curve $f(\theta)=sin(\theta), g(\theta)=cos(\theta)$ is 
a Devils staircase \cite{DS}, not 
differentiable at infinitely many points. Thus it is not 
surprising that $P(\theta|\Delta\theta)$ does not have a well defined 
limit for $\Delta\theta\to 0$.

\begin{figure}
 \centering
 \includegraphics[width=6cm]{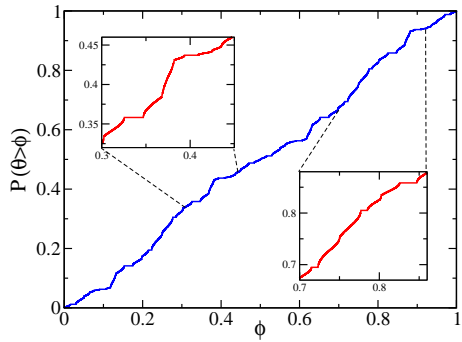}
 \caption{Cumulative distribution  $P(\theta>\phi)$  for  base curve 
$f(\theta)=\cos(\theta)$ and $g(\theta)=\sin(\theta)$  is a Devil's staircase. 
Insets show enlarged part of different regions.}
 \label{fig:cumu}
\end{figure}

\begin{figure}
 \centering
 \includegraphics[width=5.5cm]{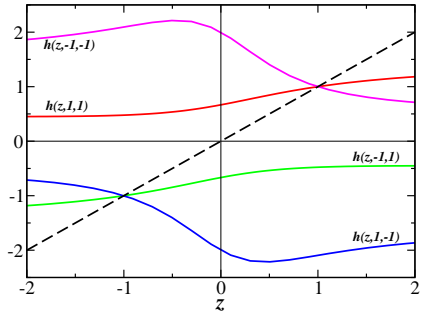}
 \caption{The stochastic non-linear map  (\ref{eq:nonlin_map}) corresponding to 
CCRW (\ref{eq:ccrw2}). The fixed points of $h(z,\sigma^x,\sigma^y)$ are $\sigma^x \sigma^y$.}
 \label{fig:fixdpt}
\end{figure}

\begin{figure}[t]
 \centering
 \includegraphics[width=6cm]{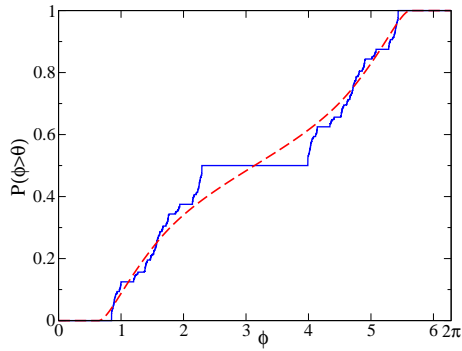}
 \caption{Cumulative distribution in the steady sate  of Eq. (\ref{eq:nonlin_map})  which 
is equivalent to CCRW (\ref{eq:ccrw2}), is compared  with  Eq. (\ref{eq:z_lin}).}
 \label{fig:cum}
\end{figure}

Contrary to naive expectation that  any deterministic dynamics, when  added 
to  an  existing  stochastic system, does not alter the the stochastic 
behavior, here we observe that  the deterministic return  of a two dimensional 
random walker to a pre-defined base curve partially destroy the random-ness 
resulting in a non-trivial  probability distribution.
To understand  it better, let us take  the following CCRW, 
\begin{eqnarray}
 {\rm Boundary } &:& |x|<1/2 ~ {\rm  and }~ |y|<1/2\cr 
 {\rm Base ~curve  ~} &:&  x= cos(\theta)/2 ~ {\rm  and }~  y = sin(\theta)/2 
\label{eq:ccrw2}
\end{eqnarray} 
where $\theta$ is angle  made with $x$-axis 
by the returning random walker. In this case,  the walker crosses the 
boundary in  every attempted walk  and returns to the base curve. 
Effectively the walker traces different points  on the base curve 
using a stochastic map $ x_{t+1}= (x_t+\sigma^x_t)/r_t ; ~ y_{t+1}= (y_t+\sigma^y_t)/r_t$ where $\sigma^{x,y}_t =\pm $ 
chosen  randomly (by the two dimensional walker during its walk) and 
$r_t = \sqrt{ (x_t+\sigma^x_t)^2 + (y_t+\sigma^y_t)}$. One can, in fact 
write an equivalent  one dimensional map in terms of $z=y/x=tan(\theta)$ as
\begin{eqnarray}
z_{t+1} = h(z_t,\sigma^x_t,\sigma^y_t)\cr~~~~ {\rm with}~~ h(z,\sigma^x,\sigma^y) =\frac{z+  2\sigma^y \sqrt{1+z^2}}{ 1+2\sigma^x\sqrt{1+z^2}}.
\label{eq:nonlin_map}
\end{eqnarray}
On each iteration $z$ takes a new value by choosing  one of the  four non-linear functions 
$h(z,\pm 1,\pm 1)$ randomly as shown in Fig. \ref{fig:fixdpt}.
These  functions $h(z,\sigma^x,\sigma^y)$ have  attractive fixed points at $z^*= \sigma^x\sigma^y$. 
Note, that if  $\sigma_x$ were not random (say  $\sigma_x=1$)  then  $z$  would evolve using
the map $z_{t+1} = h(z_t,1,\sigma^y_t)$ and ultimately remains  confined in the region $|z|<1$  
as $t\to \infty.$   Addition of the other two functions $h(z,-1,\sigma^y)$ allows  $z$ to 
move out of the region  $|z|<1$.  Corresponding distribution  of $\theta = tan^{-1} (z)$ 
(shown as a cumulative distribution in Fig. \ref{fig:cum}) compares well with  Fig. \ref{fig:cumu}, 
which indicates that CCRW  (\ref{eq:ccrw2}) has all the characteristic features of the generic  two dimensional 
bounded random walk.

We argue that this  non-trivial behavior, which occurs even for  CCRW  (\ref{eq:ccrw2}),
is the effect of the non-linearity that exists in the stochastic map (\ref{eq:nonlin_map}). 
In the following we show that   a linear stochastic map  having identical fixed points would 
not show  any such features. In particular,  the steady state probability of $z$  has a 
well-defined PDF. Let us  take the following  stochastic linear map   
\begin{equation} 
 z_{t+1} = \frac{\sigma^x_t}{2} z_t   +  (1- \frac{\sigma^x_t}{2}) \sigma^y_t
\end{equation}
which  has the same fixed points as (\ref{eq:nonlin_map}) and calculate the 
steady state distribution $g(z)$. Direct iteration of this map yields 
$$ z_{t+1} = \sum_{n=0}^{t}   \sigma^y_n (1- \frac{\sigma^x_n}{2}) 
 \frac{1}{2^{t-n}} \prod_{k=n+1}^t  \sigma^x_k.
$$
Since product of $\sigma^x$s  produces a random sign $\pm$,  we
have 
\begin{equation}
z_{t+1} = \sum_{n=0}^{t}   \frac{\tau^y_n}{2^n} (1- \frac{\tau^x_n}{2}), 
\label{eq:z_lin}
\end{equation}
where the term containing the initial value $z_0$ (being exponentially small  for 
large $t$) is dropped, and  we have taken $\tau^{x,y}_{n} = \sigma^{x,y}_{t-n}.$ 
The right hand side of the  above equation resembles  the  Hamiltonian of a   
spin system  on a lattice 
of size $L=t$, where the spins $\tau^{x,y}_n$ of two different kinds of particles $x$ and $y$
at the site $n$ interact with an inhomogeneous magnetic field $B_n= 2^{-n}.$  Thus the 
expression for the energy of the system is 
\begin{equation}
 E = \sum_{i=0}^{L}  B_i  \tau^y_i  (1- \frac{\tau^x_i}{2}). 
\label{eq:E} 
\end{equation}

The energy distribution of this model $P(E;\beta)$ in equilibrium is related to the  
stationary distribution $g(z)$ of  (\ref{eq:z_lin});   
\beq g(z) = \lim_{\beta\to 0} P(E=z;\beta) . \label{eq:Pz} \eeq
Since, for classical systems with partition function  ${\cal Z}_L(\beta)$, 
\beq
P(E,\beta) = {\cal L}^{-1} \left[  \frac{{\cal Z}_L(s+\beta)}{ {\cal Z}_L(\beta)}; E \right],\nonumber
\eeq
where  ${\cal L}^{-1}\left[f(s);x \right]=\int_\infty^\infty e^{-s x} f(s) ds$ is the  
inverse Laplace transform, we have   
\beq g(z)= {\cal L}^{-1} \left[\frac{{\cal Z}_L(s)}{ {\cal Z}_L(0)}; z\right]. \label{eq:gz}\eeq
The partition function of the model (\ref{eq:E}) 
 \begin{eqnarray}
 {\cal Z}_L(\beta) 
 &=& \frac{\sinh\beta \sinh 2\beta}{\sinh(2^{-L}\beta)\sinh(2^{-L}2\beta)}\nonumber
\end{eqnarray}
 is used  in Eq. \eqref{eq:gz} to obtain   
\beq
g(z) = \left\{ \begin{array}{ll}
 1/4 & \textrm{ $|z| <1$ }\\
 \frac {3-\epsilon(z) z }{8} & \textrm{ $1<|z|<3$}\\
 0 & \textrm{ $|z| >3$},  
  \end{array} \right.
\eeq
in thermodynamic limit  $L\to \infty$. Here $\epsilon(z)$ is the signum function. 
Correspondingly, the distribution of $\theta= \tan^{-1} z$  is $P(\theta)= g(\tan \theta) \sec^2\theta$, and its cumulative distribution
is 
   \begin{displaymath}
 P(\theta>\phi) = \left\{ \begin{array}{ll}
\frac{  2+\tan(\phi)  }{4} & \textrm{ $|tan\phi| \le 1$ }\\
 \frac  {   8+ 6 \tan\phi +\epsilon(\tan\phi) \sec\phi^2  }{16} & \textrm{ $1< |\tan \phi| <3$}\\
 0 & \textrm{ $|\tan\phi| \ge 3$}  
  \end{array} \right. .
\end{displaymath} 

In Fig. \ref{fig:cum} we have plotted the cumulative distribution $P(\theta>\phi)$ for 
both the non-linear  map  (\ref{eq:nonlin_map}) and the stochastic  linear map 
(\ref{eq:z_lin}) which has identical fixed points. It is quite evident from the figure 
that  $P(\theta>\phi)$ for the non-linear map, which corresponds to  
the CCRW (\ref{eq:ccrw2}), has a structure of Devil's staircase, whereas the 
same for the linear map is a  continuous and differentiable function.  Thus, 
it is suggestive that the non-trivial distribution  of returning walkers on the 
base curve   is an artifact of non-linearity.

 To  characterize  the nontrivial distribution $P(\theta)$ of CCRW we use  
the fractal dimension of its bounding curve.  From  naive  box-counting 
methods one expects that the bounding curve, which is a fractal, 
is covered by ${\nu}\sim \Delta \phi^{-d}$ segments of size 
$\Delta \phi$ with $d\ne 1.$ For CCRW with BC (\ref{eq:flower}) a 
plot of ${\nu}$ versus $\Delta \phi$ in log-scale (Fig. 6) shows that  $d=1.75(8)$. 
Same calculation for a few other  BCs \cite{note1} results in $d\simeq 1.75$,
which made us to conjecture that possibly the fractal dimension of PDF 
bounding curve of  CCRW  is universal ($d=7/4$).
\begin{figure}[t]
\vspace*{.5 cm}
\centering
\includegraphics[width=6cm]{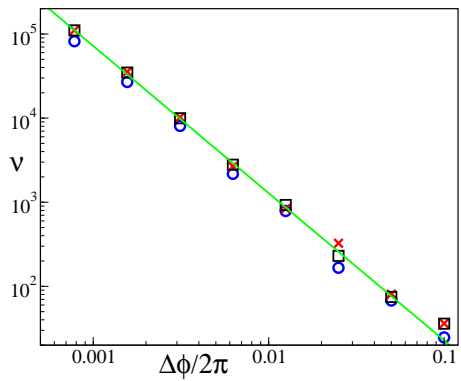}
\caption{Plot of ${\nu}$ versus  $\Delta\phi$ in log scale for the PDF of 
CCRW with BC (\ref{eq:flower}), a unit circle, and a unit square \cite{note1} are shown 
with symbols $\times$, $\circ$ and $\Box$ respectively. A line with  slope $-7/4$ 
is drawn to guide the eye.} 
\label{fig:fractald}
\end{figure} 


In conclusion, we explain that patterns of  Ciftci-Camak random walk  \cite{Ciftci}
(CCRW) are simple repetition of the base curve. An unique feature of CCRWs is that 
the distribution of returning  walkers along the base curve    
can not be represented by a functional form; corresponding cumulative 
distributions  are  Devil's staircases.  A correspondence of CCRW with stochastic Wiley
non-linear maps reveals that this unusual distribution is an artifact of non-linearity.
A quantitative  characterization of the distribution could be 
fractal dimension $d$ of its bounding curve. Our numerical calculations  of  
CCRWs with different base curves show that $d\simeq 1.75$. 

{\it Acknowledgment :} We acknowledge Urna Basu for  helpful discussions and careful reading of the manuscript.


\begin{thebibliography}{10}
\bibitem{note1}  Results for base curve i) a unit circle and ii) a unit square bounded by 
$|x|=1=|y|$ are also shown in Fig. 6.
\end{thebibliography}
\end{document}